\documentclass[a4paper,12pt]{article}
\usepackage{graphics}

\begin{document}

\title{Electroproduction of electron-positron pair
in oriented crystal at high energy}
\author{V. N. Baier
and V. M. Katkov\\
Budker Institute of Nuclear Physics,\\ Novosibirsk, 630090, Russia}

\maketitle

\begin{abstract}
The process of electroproduction of the electron-positron pair by
high energy electron in an oriented single crystal is investigated.
Two contributions are considered: the direct (one-step) process via
the virtual intermediate photon and the cascade(two-step) process
when the electron emits the real photon moving in the field of axis
and afterwards the photon converts into the pair. The spectrum of
created positron(electron) is found. It is shown that the
probability of the process is strongly enhanced comparing with the
corresponding amorphous medium.

\end{abstract}

\newpage

\section{Introduction}
The photon emission from high energy electron and electron-positron
pair creation by a photon are the basic electrodynamics processes in
oriented crystal at high energy. The probabilities of these
processes are strongly enhanced comparing with the corresponding
amorphous medium \cite{BKS}. The photon emitted by electron can
produce the electron-positron pair (the electroproduction or the
trident production process). An analysis of this process is of
evident interest as example of higher order QED process in an
oriented crystal. This process is the first step of the
electromagnetic cascade developed in a crystal by the incident
electron (the cascade processes in oriented crystals was considered
in \cite{BKS2}, see also Sec.20 in \cite{BKS}), it possesses many
important peculiarities which will be discussed below.

Recently authors developed a new approach to analysis of pair
creation by a photon \cite{BK0} and radiation from high energy
electrons \cite{BK3} in oriented crystals . This approach permits to
consider simultaneously both the coherent and incoherent mechanisms
of these processes. For calculation of the energy spectrum of
created particle in the trident production process one has to
integrate over spectrum of emitted photons. This means that both the
mentioned mechanisms (contributing in the different parts of photon
spectrum) should be taken into consideration and the developed
approach is very suitable for solution of the problem.

One has to consider two different contributions into the probability
of the process. The first one is the direct(one-step)
electroproduction of pair via the virtual intermediate photon. The
second one is the cascade(two-step) process when an electron emits
the real photon moving in the field of axis and afterwards the
photon converts into the pair. The interrelation of these
contributions depends on the target thickness $l$ since the
probability of the direct process is proportional to $l$ while the
probability of the cascade process is proportional to $l^2/2$. In
oriented crystals the real photon emission is strongly enhanced.

The electroproduction process in the oriented germanium crystal was
studied recently in the experiment NA63 at SPS at CERN (for proposal
see \cite{NA63}). The electroproduction process in an amorphous
medium was considered recently by authors \cite{BK1}.

The direct process is considered in Sec.2. The cascade process is
studied in Sec.3. The summary contribution of both processes is
analyzed using various approaches in Sec.4 including the conditions
of the NA63 experiment. It is shown in Sec.5 that the created
positron spectrum in the case of not very high the initial electron
energy $\varepsilon < 200~$GeV and for soft positrons depends very
weakly on the quite large energy loss.

\section{Direct process}

The contribution of the direct process into the probability of
electroproduction in an amorphous medium was considered in detail in
\cite{BK1}. It is shown that in a thin target this contribution can
be the essential part of the probability and dominates under some
conditions. In an oriented crystal due to the strong enhancement of
real photon emission comparing with the virtual one the contribution
of the direct process is relatively small and one can use the
equivalent photon method having a sufficient accuracy. The
equivalent photon spectrum with the crystal field taken into account
can be written within the logarithmic accuracy in the form (compare
with Eq.(7) in \cite{BK1})
\begin{equation}
n_E(y,z)=\frac{\alpha}{\pi}\frac{1-y}{y}\left[\left(1+\frac{y^2}{2(1-y)}
\right)\ln \frac{Q_E^2(y,z)}{q_E^2}-1\right],
 \label{1}
\end{equation}
where
\begin{eqnarray}
&& Q_E^2(y,z=\frac{m^2\omega^2}{\varepsilon_+\varepsilon_-}
\left(1+\frac{\varepsilon_+\varepsilon_-}{\omega^2}\kappa\right)^{2/3}=
\frac{m^2y^2}{z(y-z)}\left(1+\frac{z(y-z)}{y^2}\kappa\right)^{2/3},
\nonumber \\
&& q_E^2=\frac{m^2\omega^2}{\varepsilon(\varepsilon-\omega)}
\left(1+\frac{\varepsilon-\omega}{\omega}\chi\right)^{2/3}=
\frac{m^2y^2}{1-y}\left(1+\frac{1-y}{y}\chi\right)^{2/3},
\nonumber \\
&& \chi(\varepsilon)=\frac{eE}{m^2}\frac{\varepsilon}{m},\quad
\kappa(\omega)=\frac{eE}{m^2}\frac{\omega}{m},\quad
y=\frac{\omega}{\varepsilon},\quad
z=\frac{\varepsilon_+}{\varepsilon},\quad \kappa=y\chi,
 \label{2}
\end{eqnarray}
where $E$ is the electric field of crystal axis(plane) transverse to
the velocity vector, $\varepsilon$ is the energy of the initial
electron, $\varepsilon_+,\varepsilon_-$ are the energy of particles
of the created pair, $\omega=\varepsilon_+ + \varepsilon_-$ is the
photon energy. Here the appearance of the additional factors
depending on the parameters $\chi$ and $\kappa$ is connected with
expansion of the characteristic angles in the pair creation block
and in the equivalent photon emission block at large values of these
parameters. This item is discussed in the book \cite{BKS}, Sec.6.2
and 6.3.

In an amorphous medium  the characteristic momentum transfers are
changed due to influence of the multiple scattering on the photon
emission process (the Landau-Pomeranchuk-Migdal (LPM) effect). It is
shown in \cite{BK1} that, when one takes into account the
contribution of photons emitted by an electron at fly in a target
(boundary photons) and the change of the lower limit of momentum
transfer $q_{min}^2$ under influence of the LPM effect, the value of
$q_{min}^2$ is restored in the summary contribution:
$q_{min}^2=m^2y^2/(1-y)$. Similar situation occurs in an oriented
crystal if one takes into consideration the radiation of boundary
photons by an electron at fly in a crystal field. The spectral
distribution of boundary photons emitted by an electron at fly in a
homogeneous electric field was found in \cite{JW}, \cite{BKS1}. In
our notation this distribution with the logarithmic accuracy
\cite{JW} has the form
\begin{equation}
\frac{dw_i}{dy}=\frac{\alpha}{\pi}\frac{1-y}{y}\left(1+\frac{y^2}{2(1-y)}
\right)\ln \left(1+\frac{1-y}{y}\chi\right)^{2/3}.
\label{3}
\end{equation}
Putting Eqs.(\ref{1}) and (\ref{3}) together one obtains
\begin{eqnarray}
&&n_E(y,z)+\frac{dw_i}{dy}=\frac{\alpha}{\pi}\frac{1-y}{y}\Bigg[\left(1+\frac{y^2}{2(1-y)}
\right)\left(\ln \frac{1}{\xi}+\frac{2}{3}\ln
\left(1+z\left(1-\frac{z}{y}\right)\chi\right)\right)
\nonumber \\
&&-1\Bigg],\quad \xi=\frac{z(y-z)}{1-y}.
 \label{4}
\end{eqnarray}
In the range of applicability of the equivalent photon method ($z
\ll 1$) and for not very high energy of the initial electron when
$z\chi(\varepsilon)=\chi(\varepsilon_+) \leq 1$ the summary spectral
distribution can be presented as
\begin{equation}
\frac{dw_s}{dy}=\frac{\alpha}{\pi}\frac{1-y}{y}\left[\left(1+\frac{y^2}{2(1-y)}
\right)\ln \frac{1}{\xi}-1\right]. \label{5}
\end{equation}

\section{Cascade process}

Basing on Eqs.(16) and (17) of \cite{BK0} (see also Eq.(7.135) in
\cite{BKS}) one get the general expression for the spectral
distribution of particles of pair created by a photon
\begin{eqnarray}
&& dW(\omega, y_p)=\frac{\alpha m^2}{2\pi \omega}
\frac{dy_p}{y_p(1-y_p)} \int_0^{x_0}\frac{dx}{x_0}G(x, y_p),
\nonumber \\
&&G(x, y_p)=\int_0^{\infty} F(x, y_p, t)dt +s_3\frac{\pi}{4},
\nonumber \\
&& F(x, y_p, t)={\rm Im}\left\lbrace e^{f_1(t)}\left[s_2\nu_0^2
(1+ib)f_2(t)-s_3f_3(t) \right] \right\rbrace,\quad
b=\frac{4\kappa_1^2}{\nu_0^2}, \quad
y_p=\frac{\varepsilon_+}{\omega},
\nonumber \\
&& f_1(t)=(i-1)t+b(1+i)(f_2(t)-t),\quad
f_2(t)=\frac{\sqrt{2}}{\nu_0}\tanh\frac{\nu_0t}{\sqrt{2}},
\nonumber \\
&&f_3(t)=\frac{\sqrt{2}\nu_0}{\sinh(\sqrt{2}\nu_0t)}, \label{2c}
\end{eqnarray}
where
\begin{equation}
s_2=y_p^2+(1-y_p)^2,~s_3=2y_p(1-y_p),~\nu_0^2=4y_p(1-y_p)
\frac{\omega}{\omega_c(x)},~\kappa_1=y_p(1-y_p)\kappa(x). \label{3c}
\end{equation}

The situation is considered when the electron angle of incidence
$\vartheta_0$ (the angle between electron momentum {\bf p} and the
axis (or plane)) is small $\vartheta_0 \ll V_0/m$. The axis
potential (see Eq.(9.13) in \cite{BKS}) is taken in the form
\begin{equation}
U(x)=V_0\left[\ln\left(1+\frac{1}{x+\eta} \right)-
\ln\left(1+\frac{1}{x_0+\eta} \right) \right], \label{4c}
\end{equation}
where
\begin{equation}
x_0=\frac{1}{\pi d n_a a_s^2}, \quad  \eta_1=\frac{2
u_1^2}{a_s^2},\quad x=\frac{\varrho^2}{a_s^2}, \label{5c}
\end{equation}
Here $\varrho$ is the distance from the axis, $u_1$ is the amplitude
of thermal vibration, $d$ is the mean distance between atoms forming
the axis, $a_s$ is the effective screening radius of the potential.
The parameters in Eq.(\ref{4c}) were determined by means of fitting
procedure, see Table 1.

The local value of the parameter $\kappa(x)$,  which determines the
probability of pair creation in the field Eq.(\ref{4c}), is
\begin{equation}
\kappa(x)=-\frac{dU(\varrho)}{d\varrho}\frac{\omega}{m^3}=2\kappa_sf(x,\eta),\quad
f(x,\eta)=\frac{\sqrt{x}}{(x+\eta)(x+\eta+1)},\quad
\kappa_s=\frac{V_0 \omega}{m^3a_s}\equiv \frac{\omega}{\omega_s}.
 \label{6c}
\end{equation}
For an axial orientation of crystal the ratio of the atom density
$n(\varrho)$ in the vicinity of an axis to the mean atom density
$n_a$ is (see \cite{BK0})
\begin{equation}
\frac{n(x)}{n_a}=\xi(x)=\frac{x_0}{\eta_1}e^{-x/\eta_1},\quad
\omega_0=\frac{\omega_e}{\xi(0)}, \quad
\omega_e=4\varepsilon_e=\frac{m}{4\pi
Z^2\alpha^2\lambda_c^3n_aL_0}.\label{7c}
\end{equation}

The functions and values in Eqs.(\ref{2c}) and (\ref{3c}) are (see
\cite{BK2})
\begin{eqnarray}
&&\omega_c(x)=
\frac{\omega_e(n_a)}{\xi(x)g_p(x)}=\frac{\omega_0}{g_p(x)}e^{x/\eta_1},\quad
L=L_0g_p(x),  \quad L_0=\ln(ma)+ \frac{1}{2}-f(Z\alpha),
\nonumber \\
&&
 g_p(x)=g_{p0}+\frac{1}{6 L_0}\left[\ln
\left(1+\kappa_1^2\right)+\frac{6 D_{p}\kappa_1^2}
{12+\kappa_1^2}\right],\quad
g_{p0}=1-\frac{1}{L_0}\left[\frac{1}{42}+h\left(\frac{u_1^2}{a^2}\right)\right],
\nonumber \\
&&  h(z)=-\frac{1}{2}\left[1+(1+z)e^{z}{\rm Ei}(-z) \right], \quad
a=\frac{111Z^{-1/3}}{m}. \label{8c}
\end{eqnarray}
Here the function $g_p(x)$ determines the effective logarithm using
the interpolation procedure, $D_{p}=D_{sc}-10/21=1.8246$,
$D_{sc}=2.3008$ is the constant entering in the radiation spectrum
at $\chi/u \gg 1$ (or in the positron spectrum in the pair creation
process at $\kappa_1 \gg 1$), see Eq.(7.107) in \cite{BKS},~ Ei($z$)
is the integral exponential function.

The expression for the spectral probability of radiation is
connected with the spectral distribution Eq.(\ref{2c})
($dW/dy=\omega dW/d\varepsilon $) by the standard QED substitution
rules: $\varepsilon_+ \rightarrow -\varepsilon,~\omega \rightarrow
-\omega,~\varepsilon_+^2d\varepsilon_+ \rightarrow \omega^2d\omega$
and exchange $\omega_c(x) \rightarrow 4\varepsilon_c(x)$. As a
result one has for the spectral intensity $dI=\omega dW$
\begin{eqnarray}
&& dW_r(\varepsilon,y)=\frac{\alpha m^2}{2\pi \varepsilon}
\frac{\omega dy}{1-y} \int\limits_0^{x_0}\frac{dx}{x_0}G_{r}(x, y),
\nonumber \\
&&G_{r}(x, y)=\int\limits_0^{\infty} F_{r}(x, y, t)dt
-r_{3}\frac{\pi}{4},
\nonumber \\
&& F_{r}(x, y, t)={\rm Im}\left\lbrace
e^{\varphi_1(t)}\left[r_{2}\nu_{0r}^2 (1+ib_r)f_2(t)+r_{3}f_3(t)
\right] \right\rbrace,\quad b_r=\frac{4\chi^2(x)}{u^2\nu_{0r}^2},
\nonumber \\
&& y=\frac{\omega}{\varepsilon}, \quad u=\frac{y}{1-y},\quad
\varphi_1(t)=(i-1)t+b_r(1+i)(f_2(t)-t), \label{r1}
\end{eqnarray}
where
\begin{eqnarray}
&&r_2=1+(1-y)^2,\quad r_3=2(1-y),\
\nonumber \\
&&\nu_{0r}^2=\frac{1-y}{y} \frac{\varepsilon}{\varepsilon_c(x)},
\label{r2}
\end{eqnarray}
here the functions $f_2(t)$ and $f_3(t)$ are defined in
Eq.(\ref{2c}). The local value of the parameter $\chi(x)$  which
determines the radiation probability in the field Eq.(\ref{4c}) is
\begin{equation}
\chi(x)=-\frac{dU(\varrho)}{d\varrho}\frac{\varepsilon}{m^3}=2\chi_s
f(x,\eta),\quad \chi_s=\frac{V_0 \varepsilon}{m^3a_s}\equiv
\frac{\varepsilon}{\varepsilon_s},
 \label{r3}
\end{equation}
where $f(x)$ is defined in Eq.(\ref{6c}).

The functions and values in Eqs.(\ref{r1}) and (\ref{r2}) (see also
Eqs.(\ref{7c}) and (\ref{8c})) are
\begin{eqnarray}
&&\varepsilon_c(x)=
\frac{\varepsilon_e(n_a)}{\xi(x)g_r(x)}=\frac{\varepsilon_0}{g_r(x)}e^{x/\eta_1},
\nonumber \\
&&g_r(x)=g_{r0}+\frac{1}{6 L_0}\left[\ln
\left(1+\frac{\chi^2(x)}{u^2}\right)+\frac{6 D_{r}\chi^2(x)}
{12u^2+\chi^2(x)}\right],
\nonumber \\
&&
g_{r0}=1+\frac{1}{L_0}\left[\frac{1}{18}-h\left(\frac{u_1^2}{a^2}\right)\right],\quad
\label{r4}
\end{eqnarray}
where the function $g_r(x)$ determines the effective logarithm using
the interpolation procedure:$L=L_0g_r(x)$, see Eq.(\ref{8c}),
$D_r=D_{sc}-5/9$=1.7452.

The expressions for the radiation probability $dW_r$ Eq.(\ref{r1})
and the pair creation probability $dW$ Eq.(\ref{2c}) include both
the coherent and incoherent contributions as well as the influence
of the multiple scattering (the LPM effect) on the photon emission
process (see \cite{BK2}) or the pair creation process (see
\cite{BK3}) . The probability of the coherent radiation is the first
term ($\nu_{0r}^2=0$) of the decomposition of Eq.(\ref{r1}) over
$\nu_{0r}^2$ and the probability of the coherent pair creation is
the first term ($\nu_{0}^2=0$) of the decomposition of Eq.(\ref{2c})
over $\nu_{0}^2$. The probabilities of the incoherent process are
the second term ($\propto \nu_0^2, \nu_{0r}^2$) of the mentioned
decompositions (compare with Eq.(21.21) in \cite{BKS}).

In the cascade process the initial electron with the energy
$\varepsilon$ emitted the photon with the energy $\omega$ which
creates in turn the electron-positron pair with the energies
$\varepsilon_+,\varepsilon_-, \omega=\varepsilon_+ +\varepsilon_-$.
The probability of the cascade process (in 1/cm$^2$) is
\begin{equation}
\frac{dW_c(z,\varepsilon)}{dz}=\int\limits_z^1
\frac{dW_r(\varepsilon,y)}{dy}\frac{dW\left(y\varepsilon ,
\frac{z}{y}\right)}{dz}dy,
 \label{c1}
\end{equation}
where $dW_r(\varepsilon,y)/dy$ is given in Eq.(\ref{r1}) and
$dW(\omega,y_p)/dz$ is given in Eq.(\ref{2c}), $\omega=
y\varepsilon,~ y_p=z/y$.

It is known that the intensity of radiation in an oriented crystal
is strongly amplified comparing with the corresponding amorphous
medium, e.g. in the oriented along axis $<110>$ germanium crystal at
the temperature T=293 K the radiation length at electron energy
$\varepsilon=$180 GeV is 22 times shorter than in the amorphous
germanium, see Sec.17 in \cite{BKS}. The spectra of photons emitted
from electrons with different energies in the oriented germanium are
shown in Fig.1. The spectra have the wide maximum situated at
$y=0.05$ for $\varepsilon=126~$GeV (the corresponding photon energy
is $\omega_h=6.3~$GeV), at $y=0.06$ for $\varepsilon=153~$GeV (the
corresponding photon energy is $\omega_h=9.2~$GeV), and at $y=0.07$
for $\varepsilon=180~$GeV (the corresponding photon energy is
$\omega_h=12.6~$GeV), while for $\varepsilon=500~$GeV the maximum is
situated at $y=0.12$ and the corresponding photon energy is
$\omega_h=60~$GeV. The magnitudes of maxima for three first cases
are very close to each other (the difference is less than 2\%).

The emitted photons create the electron-positron pairs. Since the
spectrum of created positron has overall drop $\sim 1/z$ it is
instructive to plot the function $zdW_c/dz$ to trace details of the
pair creation mechanism as it was done in \cite{BK1}. The values of
the function in the oriented along axis $<110>$ germanium crystal at
temperature T=293 K are shown in Fig.2 for two energies
$\varepsilon$=180~GeV and 500~GeV. There are two mechanisms of pair
creation by a photon in an oriented crystal: the coherent and the
incoherent. When $\kappa_m \ll 1$
($\kappa_m=\omega/\omega_m,~\omega_m$ is contained in the Table 1)
the incoherent mechanism of pair creation (weakly depending on the
photon energy) dominates. Just this case exhibits the curve 1 and
the left part of curve 2 in Fig.2. Starting from $\omega \simeq
2\omega_m/3$  the coherent mechanism comes into play (see e.g. the
second term in Eq.(\ref{c1a}) below, see also Fig.1 in \cite{BK1}).
Note that $2\omega_m/3 \simeq 60~$GeV is the position of the maximum
at $\varepsilon=500~$ GeV in Fig.1.  So the the large peak at the
right part of curve 2 in Fig.2 is the coherent contribution.

The radiation spectra in the oriented along axis $<111>$ tungsten
crystal at temperature T=293 K are shown in Fig.3 of \cite{BK1}. The
spectra have the wide maximum situated at $y=0.003$ for
$\varepsilon=1~$GeV (the corresponding photon energy is
$\omega_h=3~$MeV), at $y=0.009$ for $\varepsilon=3~$GeV (the
corresponding photon energy is $\omega_h=27~$MeV), at $y=0.01$ for
$\varepsilon=5~$GeV (the corresponding photon energy is
$\omega_h=50~$MeV), and at $y=0.02$ for $\varepsilon=10~$GeV (the
corresponding photon energy is $\omega_h=200~$MeV), The magnitudes
of maxima for all four cases are very close to each other (the
maximal difference is $\sim 2.5$\%). For higher electron energies
the maximum is shifted to the right: for temperature T=100 K it is
at $y=0.08$ for $\varepsilon=50~$GeV (the corresponding photon
energy is $\omega_h=4~$GeV) and at $y=0.1$ for $\varepsilon=100~$GeV
(the corresponding photon energy is $\omega_h=10~$GeV)

The functions $zdW_c/dz$ for cascade pair electroproduction in the
indicated conditions in tungsten are shown in Fig.3 for three
energies $\varepsilon$=10~GeV, 50~GeV and 100~GeV. Since for
$\varepsilon$=10~GeV the maximum position $\omega_h
\ll\omega_m\simeq 8~$GeV the incoherent mechanism of pair creation
(weakly depending on the photon energy) dominates (the curve 1). The
same remains true for the soft positron production at higher
electron energies.  The peaks at the right part of positron spectra
for $\varepsilon$=100~GeV and $\varepsilon$=50~GeV situated at
$z=0.2$ are the contributions of the coherent radiation at the first
stage of the process. For $\varepsilon$=100~GeV the maximum position
$\omega_h=10~$GeV is twice of higher than 2$\omega_m/3 \simeq~$5 GeV
and the region of spectrum maximum contributes significantly (the
curve 3), while for $\varepsilon$=50~GeV $\omega_h=4~$GeV is
slightly less than 2$\omega_m/3$ and the right part of the photon
spectrum contributes only (the curve 2). This explains the sharp
difference in heights of the peaks.

\section{Summary probability of electroproduction }

In the experiment NA63 carried out recently at SPS at CERN (for
proposal see \cite{NA63}) the electroproduction (trident production)
process was studied in the oriented along axis $<110>$ germanium
crystal at temperature T=293 K. Targets with the thickness
$l=170~\mu m$ and $l=400~\mu m$ were used. The theory prediction in
the energy interval measured in NA63 is shown in Fig.4. The direct
process contribution (in 1/cm) is
\begin{equation}
\frac{dW_d(z,\varepsilon)}{dz}=\int\limits_z^1
\frac{dw_{s}(\varepsilon,y)}{dy}\frac{dW\left(y\varepsilon ,
\frac{z}{y}\right)}{dz}dy,
 \label{c2}
\end{equation}
where $dw_{s}(z,\varepsilon)/dy$ is given in Eq.(\ref{5}). The
cascade process contribution (in 1/cm$^2$)is defined in
Eq.(\ref{c1}).

The summary contribution into the pair electroproduction probability
of both the direct and the cascade mechanisms in an oriented crystal
in the target with thickness $l$ is
\begin{equation}
\frac{dW_S(z,\varepsilon)}{dz}=\frac{dW_d(z,\varepsilon)}{dz} l+
\frac{dW_c(z,\varepsilon)}{dz}\frac{l^2}{2}.
 \label{c3}
\end{equation}
The curve 1 in Fig.4 represents this summary contribution for
$l=400~\mu m$ while the curve 3 shows the same for $l=170~\mu m$.
The relative contribution of the direct process is maximal at the
minimal positron energy $\varepsilon=0.5$~GeV. For $l=170~\mu m$ the
direct process contribution is $\sim 22\%$ of the cascade one while
it is  only $\sim 9\%$ for $l=400~\mu m$. This relative contribution
diminishes with the positron energy increase and at the positron
energy $\varepsilon=9$~GeV it becomes $\sim 7\%$ of the cascade
contribution for $l=170~\mu m$ and only $\sim 3\%$  for $l=400~\mu
m$.

It is customary to present the result in terms of the enhancement:
the ratio of the electroproduction probability in an oriented
crystal Eq.(\ref{c3}) to the corresponding probability in an
amorphous medium \cite{BK1}. The enhancement for two used
thicknesses is shown in Fig.5. The increase of the enhancement with
the positron energy growth for the given thickness is due to more
fast decreasing of the probability in an amorphous medium. In the
positron energy interval under consideration in an amorphous medium
the main contribution into electroproduction probability gives
direct process $\propto l$ (see Fig.2 in \cite{BK1}), while in
oriented crystal the cascade process probability $\propto l^2$
dominates. Because of this one has the higher enhancement for
$l=400~\mu m$.

We will discuss now an approximate approach for the consideration of
pair electroproduction in the cascade process. The probability of
the cascade process Eq.(\ref{c1}) depends on the probabilities of
pair creation by a photon and photon emission from high-energy
electron. The photon spectrum has the maximum in its soft part (e.g.
in Ge $\omega_h=12.6~$GeV for $\varepsilon=180~$GeV). From the other
side it is known that for the pair creation process in germanium
crystal the coherent (field) contribution becomes essential starting
with the energy $\varepsilon=50~$GeV (see Fig.2a in \cite{BK2}). So
one can expect that for $\varepsilon_+ \leq \varepsilon_m$ the main
contribution in the pair creation part $dW(y\varepsilon, z/y)/dy$
gives the incoherent pair creation probability with  the field
(coherent) correction:
\begin{eqnarray}
&&\frac{dW\left(y\varepsilon, \frac{z}{y}\right)}{dz}\rightarrow
\frac{dW_{ic}(\varepsilon, z)}{dz}=\frac{g_{p0}}{yL_{rad}}\Bigg[
\left(1-\frac{4z(y-z)}{3y^2}\right)
\nonumber \\
&&+B
\left(1-\frac{z(y-z)}{y^2}\right)\exp\left(-\frac{2\varepsilon_m
y}{3\varepsilon z(y-z)}\right)\Bigg],
\nonumber \\
&& B=\frac{\pi \omega_0}{\eta_1\omega_m
g_{p0}}\sqrt{-\frac{3f(x_m)}{4f''(x_m)}},\quad
x_m=\frac{1}{6}\left(\sqrt{1+16\eta(1+\eta)}-1-2\eta\right),
 \label{c1a}
\end{eqnarray}
where $L_{rad}$ is the Bethe-Maximon radiation length, see e.g.
Eq.(7.54) in \cite{BKS}, and Eqs.(21.29) in the same book, the
function $f(x)$ is defined in Eq.(\ref{6c}). Here the first term in
r.h.s. is the Bethe-Maximon probability with crystal effects taken
into account ($g_{p0}$ is defined in Eq.(\ref{8c})) and the second
term is the field correction calculated according to Eq.(12.14) in
\cite{BKS} which is valid near the threshold of pair creation by a
photon in the field. For considered cases in germanium B=24.5 and in
tungsten B=6.16.  The result of calculation of the cascade
electroproduction probability is given in Fig.6. The curve 1 is the
probability of the process according to Eq.(\ref{c1}) and the curve
2 is for the case when the pair creation probability is used in form
Eq.(\ref{c1a}). The maximal difference between the curves is 5\%
only (at $z=0.008$). It should be noted that the coherent pair
creation contribution (the second term in Eq.(\ref{c1a})) gives
negligible contribution at $z \leq 0.05$, gives 7\% of total
contribution at $z=0.1$ and is nearly half of total contribution at
$z=0.5$.

The main contribution into the spectrum of low energy positrons $z
\ll 1$ gives soft photons $y \sim z \ll 1$. If the condition
$x_0^{-3/2} \ll y/\chi_s \ll 1$ is fulfilled one can use
Eqs.(17.11)-(17.13) in \cite{BKS}) for the description of the
spectrum of emitted photons. The approximate spectral intensity
distribution can be presented in the form
\begin{equation}
y\frac{dW_{a}(y)}{dy}=\frac{dI_a}{d\omega}=\frac{A}{L_{rad}}
\left(\frac{y}{\chi_s}\right)^{1/3}g_1(y,\eta),
 \label{c4}
\end{equation}
where
\begin{eqnarray}
&&A=\left(\frac{2}{\sqrt{3}}\right)^{5/3}
\Gamma\left(\frac{2}{3}\right)\frac{\omega_0}{\eta_1\varepsilon_s},\quad
g_1(y,\eta)=\ln\frac{\chi_s}{y}+a(\eta),
\nonumber \\
&&a(\eta)=\ln(18\sqrt{3})-\frac{\pi}{2\sqrt{3}}-C-\frac{3}{4}-
l_1(\eta), \quad C=0.577...,\quad
\nonumber \\
&& l_1(\eta)=\frac{3}{2}\int\limits_0^{\infty}\left(f^{2/3}(x,
0)-f^{2/3}(x, \eta)\right)dx, \label{c5}
\end{eqnarray}
Here the function $f(x, \eta)$ is defined in Eq.(\ref{6c}), the
parameter $\chi_s$ is defined in Eq.(\ref{r3}).

The spectral intensity Eq.(\ref{c4}), which describes the radiation
in the crystal field (the coherent radiation), is in quite
satisfactory agreement with the spectral curves in Fig.1 not far
from maximum. The incoherent contribution in the considered photon
energy interval are damped as
$(y/\chi_m)^{2/3}~(\chi_m=\varepsilon/\varepsilon_m)$ comparing with
the amorphous medium (see Eq.(21.23) in \cite{BKS})).  The LPM
effect is damped more stronger. In the integral over the variable
$x$, which defined the function $g_1(y,\eta)$, the large values $x$
contributed up to $x \sim \chi_s/y$. For very soft photons $y \leq
\chi_s/x_0^{3/2}$ all the interval $0\leq x \leq x_0$ contributes.
In the limiting case $y \ll \chi_s/x_0^{3/2}$ one has
\begin{equation}
y\frac{dW_{a}(y)}{dy}=\frac{dI_a}{d\omega}=\frac{A}{L_{rad}}
\left(\frac{y}{\chi_s}\right)^{1/3}g_2(y,\eta),
 \label{c6}
\end{equation}
where
\begin{eqnarray}
&& g_2(y,\eta)=\frac{3}{2}\int\limits_0^{x_0}f^{2/3}(x, \eta)dx
\simeq \frac{3}{2}\int\limits_0^{x_0}f^{2/3}(x, 0)dx-l_1(\eta)
\nonumber \\
&& \simeq \frac{3}{2}\ln x_0+\frac{9}{4} \ln 3
-\frac{\pi\sqrt{3}}{4} -l_1(\eta). \label{c7}
\end{eqnarray}

One can calculate the probability of the electroproduction process
substituting the probabilities Eq.(\ref{c4}) and Eq.(\ref{c1a}) into
Eq.(\ref{c1}). For $z \leq 0.1$ the obtained probability is
20\%-30\% higher than the direct calculation of Eq.(\ref{c1}). This
is due to the fact that the whole spectrum of emitted photons (not
only the vicinity of the maximum) is contributed into the final
electroproduction probability and the tails of the curve
Eq.(\ref{c4}) are higher than the spectrum Eq.(\ref{r1}).

The exact calculation of Eq.(\ref{c1}), which is the 5-fold
integral, is quite cumbersome. We used the interpolated photon
spectrum (the accuracy of interpolation is better than 1\%) to
simplify the calculation. When one uses the explicit expression
Eq.(\ref{c1a}) for the pair photoproduction probability, one can
calculate the 3-fold integral directly. For the initial electron
energy $\varepsilon < 200~$Gev the accuracy of the result is quite
satisfactory.   The simplest calculation (the 1-fold integral) is
with use of the probabilities Eq.(\ref{c4}) and Eq.(\ref{c1a}), but
the result can be considered as a rough approximation only.

\section{Conclusion}

The process of the electroproduction of electron-positron pair (the
trident production) in an oriented crystal is considered for the
first time. It is shown that due to the strong enhancement of photon
emission in an oriented crystal (this is the coherent radiation, see
Fig.1) the electroproduction probability is also enhanced (see
Fig.5). The scale of the enhancement in the soft part of created
particles spectrum ($z \ll 1$) is similar to the photon emission
enhancement. This is connected with the fact that in this part of
the spectrum the standard incoherent (Bethe-Maximon) mechanism of
pair creation by a photon (which is weakly dependent on photon
energy) dominates (see e.g. Eq.(\ref{c1a})). For very high energy of
incident electron ($\varepsilon > \varepsilon_m$) there is an
additional enhancement in the hard part of created particles
spectrum due to the coherent pair creation mechanism (see curve 2 in
Fig.2). Indeed the enhancement $E_c$ of the cascade process (the
ratio of the electroproduction probability in the germanium crystal,
axis $<110>$, T=293 K and of the electroproduction probability in
the amorphous germanium) for the initial electron energy
$\varepsilon=500~$GeV attains $E_c \simeq 430$ at $z=0.35$. This is
because the both factors of the enhancement are acting: the coherent
radiation $\sim 20$ and the coherent pair creation $\sim 20$.

The similar situation occurs in the tungsten crystal. All the curves
in Fig.3 are close to each other at $z \simeq 0.02$. At this value
of $z$ the enhancement of the cascade process  $E_c \simeq 20$. For
peak values at $z=0.2$ one has $E_c \simeq 100$ at
$\varepsilon=100~$GeV. This is because the both factors of the
enhancement are acting: the coherent radiation $\sim 10$ and the
coherent pair creation $\sim 10$.  From the other side  $E_c \simeq
54$ at $\varepsilon=50~$GeV. This is because the coherent pair
creation is not acting entirely at this electron energy as it was
explained above.

In the germanium crystal (axis $<110>$, T=293 K) due to the action
of the coherent mechanism of photon emission the crystal radiation
length is ${\rm L}_{ch}=1.02~$mm for the  energy
$\varepsilon=180~$GeV. So the targets used in CERN experiment
consist 40\% and 17\% of the crystal radiation length ${\rm L}_{ch}$
and the energy loss of the incident electron should be analyzed.
Using the approach developed in \cite{BK3} one can calculate the
energy loss. For the initial energy $\varepsilon=180~$GeV one has
the final energy $\varepsilon_f=153~$GeV for the target thickness
$l=170~\mu m$ and for the target thickness $l=400~\mu m$ the final
energy $\varepsilon_f=126~$GeV. So the energy loss is very
essential. However, the radiation spectra for these electron
energies (shown in Fig.1) are very close to each other as it was
indicated above. Since for electron energy $\varepsilon \leq
200~$GeV and for $z \ll 1$ the incoherent pair creation gives the
main contribution, one can expect that the influence of the energy
loss on the created positron spectrum will be quite weak. The
positron spectra for corresponding electron energies are shown in
Fig.7. The maximal difference (for the positron energy
$\varepsilon_+$=0.5~ GeV) is 8.6\%. Taking into account the weak
dependence of the positron spectrum on the initial electron energy
one can use the average energy
$\varepsilon_a=(\varepsilon+\varepsilon_f)/2$ for estimation of
electron energy inside the targets. For the target thickness
$l=400~\mu m$ one has $\varepsilon_a=153~$GeV and the influence of
the energy loss will be $\sim 4$\% for this thickness at
$\varepsilon_+$=0.5~ GeV and smaller (up to 0) at the higher
positron energies ($\varepsilon_+\leq$10~GeV ). For the target
thickness $l=170~\mu m$ the influence will be $\sim 2$\% at
$\varepsilon_+$=0.5~ GeV and correspondingly smaller at higher
positron energies. So we arrive to the paradoxical conclusion: in
spite of essential energy loss by the initial electron in the
targets the created positron spectrum depends very weakly on the
energy loss in the case of not very high electron energy
$\varepsilon < 200~$GeV and $z \ll 1$.

\vspace{0.5cm}

{\bf Acknowledgments}

The authors are indebted to the Russian Foundation for Basic
Research supported in part this research by Grant 06-02-16226.

\newpage

{\bf Figure captions}

{\bf Fig.1}~  The radiation spectral intensity vs the photon energy
$y= \omega/\varepsilon$ in germanium, axis $<110>$, temperature
T=293 K. The intensity distribution  $dI(\varepsilon, y_r)/d\omega$
(in units ${\rm cm}^{-1}$) The solid curve(1) is the theory
prediction (see Eq.(\ref{r1})) for the electron energy
$\varepsilon=$126~GeV, the dash-dot curve(2) is for the electron
energy $\varepsilon=$153~GeV, the dashed curve(3) is for the
electron energy $\varepsilon=$180~GeV, and the dotted curve(4) is
for the electron energy $\varepsilon=$500~GeV.

{\bf Fig.2}~The value $zdW_c(z,\varepsilon)/dz$ for the cascade pair
electroproduction Eq.(\ref{c1}) in germanium, axis $<110>$,
temperature T=293 K. The curves 1 and 2 are for the initial electron
energy $\varepsilon$=180~GeV and  500~GeV.

{\bf Fig.3}~The value $zdW_c(z,\varepsilon)/dz$ for the cascade pair
electroproduction Eq.(\ref{c1}) in tungsten, axis $<111>$,
temperature T=100 K. The curves 1,2 and 3 are for the initial
electron energy $\varepsilon$=10~GeV, 50~GeV  and  500~GeV.

{\bf Fig.4}~ The value $zdW_S/dz$ (see Eq.(\ref{c3})) for the pair
electroproduction in germanium. The curves 1 and 3 present the
summary contribution into the pair electroproduction probability of
both the direct and the cascade mechanisms in the oriented crystal
(axis $<110>$,~T= 293 K) and the curves 2 and 4 show the summary
contribution into the pair electroproduction probability of both the
two-photon diagrams and the cascade process in an amorphous medium
(see \cite{BK1})) for $l=400~\mu m$ and $l=170~\mu m$, respectively.
For convenience the ordinate is multiplied by $10^3$.

{\bf Fig.5}~ The enhancement: the ratio of electroproduction
probability in the germanium crystal axis $<110>$, T=293 K and the
electroproduction probability in the amorphous germanium for the
initial electron energy $\varepsilon=180~$GeV. The curve 1 is for
the target thickness $l=400~\mu m$ and the curve 2 is for the target
thickness $l=170~\mu m$.

{\bf Fig.6}~ The value $zdW_c/dz$ for the pair electroproduction in
germanium. The curve 1 is calculated according to Eqs.(\ref{c3}),
(\ref{c1}) and (\ref{c2}), while the curve 2 is calculated according
to Eq.(\ref{c3}) and Eq.(\ref{c1a}).

{\bf Fig.7}~ The value $zdW_c/dz$ for the pair electroproduction in
germanium. The curve 1 is for the initial electron energy
$\varepsilon=126~$GeV, the curve 2 is for $\varepsilon=153~$GeV, the
curve 3 is for $\varepsilon=180~$GeV.

\newpage
\begin{table}
\begin{center}
{\sc Table 1}~ {Parameters of the pair photoproduction and radiation
processes in the tungsten crystal, axis $<111>$ and the germanium
crystal, axis $<110>$ for two temperatures T
($\varepsilon_0=\omega_0/4, \varepsilon_m=\omega_m,
\varepsilon_s=\omega_s$)}
\end{center}
\begin{center}
\begin{tabular}{*{10}{|c}|}
\hline Crystal& T(K)&$V_0$(eV)&$x_0$&$\eta_1$&$\eta$&
$\omega_0$(GeV)&$\varepsilon_m$(GeV)&$\varepsilon_s$(GeV)&$h$ \\
\hline W & 293&417&39.7&0.108&0.115&29.7&14.35&34.8&0.348\\
\hline W &100&355&35.7&0.0401&0.0313&12.25&8.10&43.1&0.612\\
\hline Ge & 293 & 110& 15.5
&0.125&0.119&592&88.4&210&0.235\\
\hline Ge & 100 & 114.5& 19.8
&0.064&0.0633&236&50.5&179&0.459\\
\hline
\end{tabular}
\end{center}
\end{table}


\newpage

\begin{thebibliography}{99}

\bibitem{BKS} V. N. Baier, V. M. Katkov and V. M. Strakhovenko,
{\em Electromagnetic Processes at High Energies in Oriented Single
Crystals}, World Scientific Publishing Co, Singapore, 1998.
\bibitem{BKS2} V. N. Baier, V. M. Katkov and V. M. Strakhovenko,
Nucl.Instr.and Meth,~ B 27 (1987) 360.
\bibitem{BK0} V. N. Baier, and V. M. Katkov,
Phys.Lett.,  A 346 (2005) 359.
\bibitem{BK3} V. N. Baier, and V. M. Katkov,
Phys.Lett.,  A 353 (2006) 91.
\bibitem{BK1} V. N. Baier and V. M. Katkov, Pis'ma v ZhETP,
{\bf 88} (2008) 88.
\bibitem{JW} M. Jacob, T, T, Wu, Phys.Lett. B 197 (1987) 253.
\bibitem{BKS1} V. N. Baier, V. M. Katkov and V. M. Strakhovenko,
Nucl. Phys. B 328 (1989) 387.
\bibitem{BK2} V. N. Baier, and V. M. Katkov,
Phys.Lett.,  A 372 (2008) 2904.
\bibitem{NA63} J. U. Andersen, K.Kirsebom, S. P. Moller {\em et al},
{\em Electromagnetic Processes in Strong Cristalline Fields},
CERN-SPSC-2005-030.


\end{thebibliography}
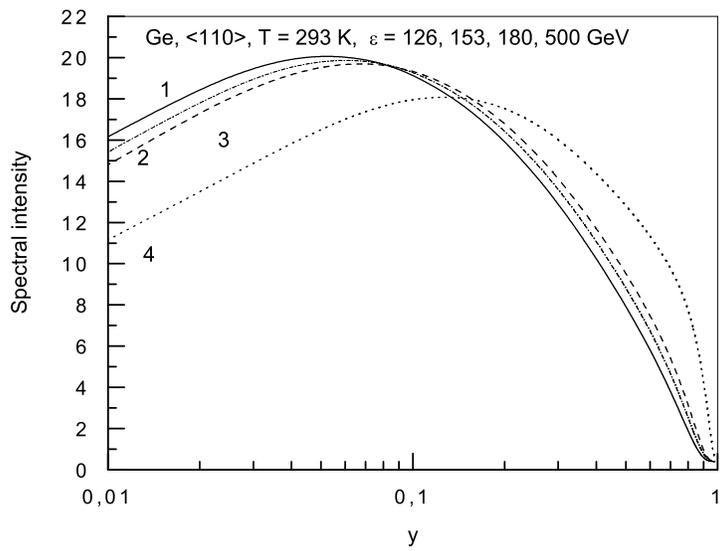
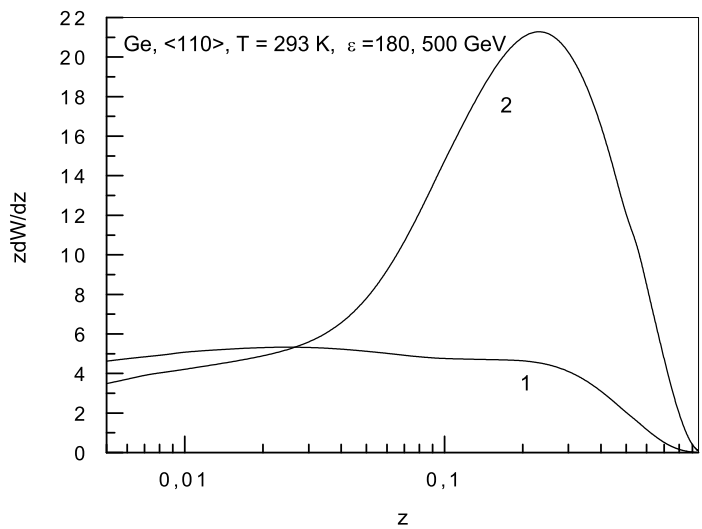
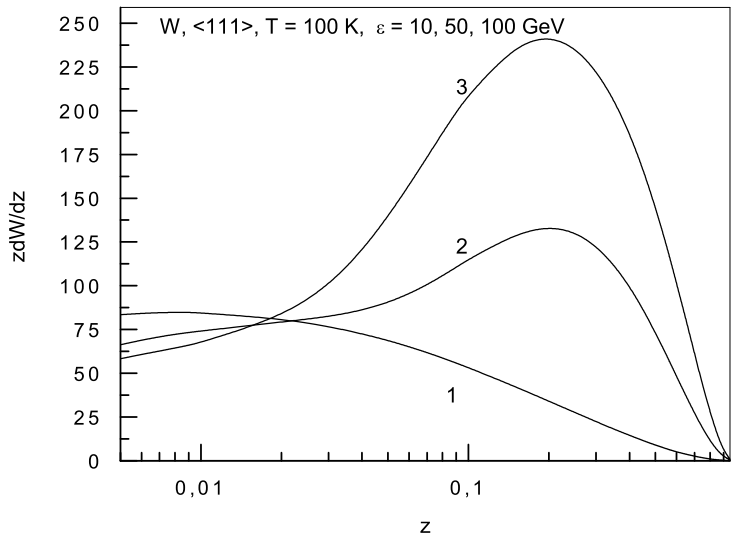
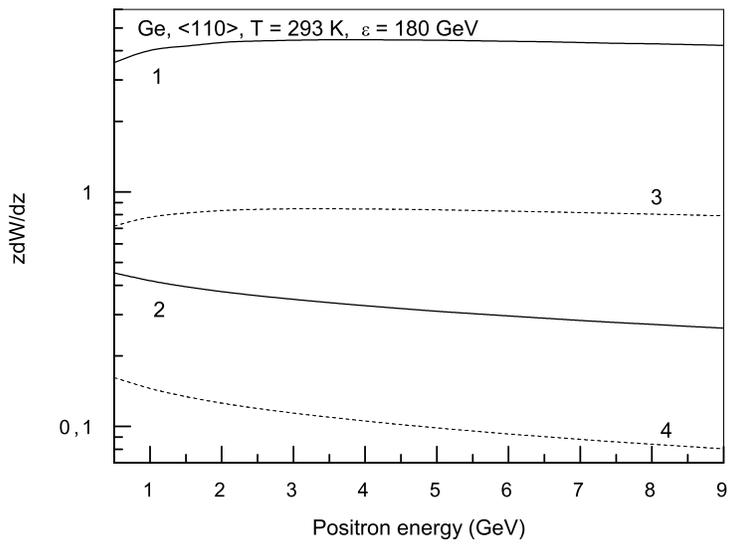
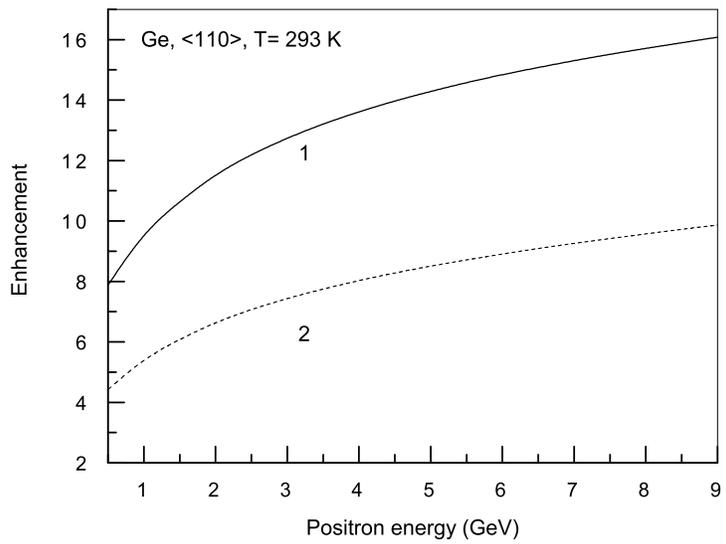
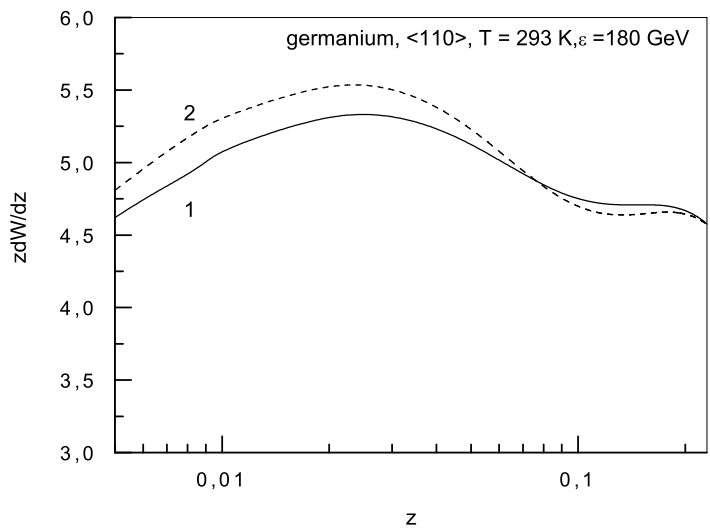
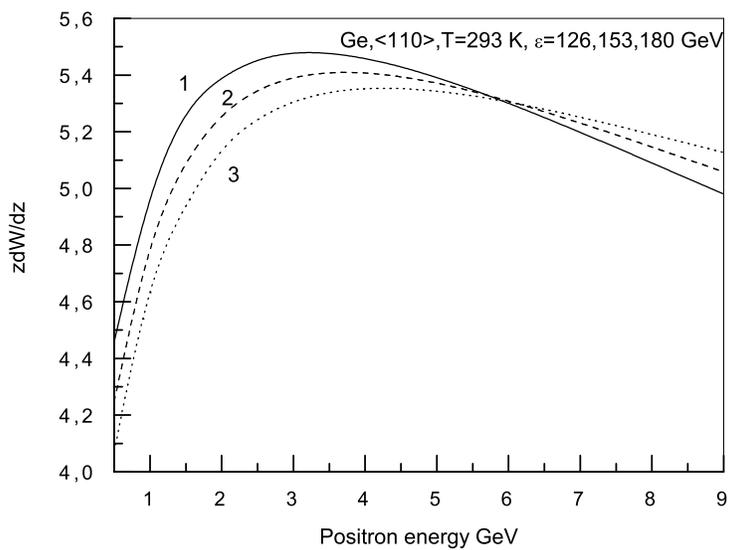
\end{document}